\documentstyle[aps,multicol]{revtex}
\input{psfig}
\begin{document}
\draft
\title{Effect of Interactions on the Admittance of Ballistic Wires}
\author{Ya.~M.~Blanter and M.~B\"uttiker}
\address{D\'epartement de Physique Th\'eorique, 
Universit\'e de Gen\`eve, CH-1211, Gen\`eve 4, Switzerland.}
\date{\today}
\maketitle 
\tighten

\begin{abstract}
A self-consistent theory of the admittance of a perfect ballistic,
locally charge neutral wire is proposed. Compared to a non-interacting
theory, screening effects drastically change the frequency behavior of
the conductance. In the single-channel case the frequency dependence
of the admittance is monotonic, while for two or more channels
collective interchannel excitations lead to resonant structures in the
admittance. The imaginary part of the admittance is typically
positive, but can become negative near resonances. 
\end{abstract}

\begin{multicols}{2}
The ac conductance (admittance) in mesoscopic systems
attracted recently strong interest, mostly due to the
finite-frequency measurements of Aharonov-Bohm oscillations in rings
\cite{Pieper} and noise in diffusive metallic wires \cite{Prober}.
The theoretical investigation of the problem raises an
important question: Standard calculations of the conductance,
employing either a scattering (Landauer) or a linear response (Kubo)
approach, describe the current in response to the {\em external}
electric field, assuming the latter to be uniform (linear response) or
ignoring the actual distribution of the potential (dc scattering
approach). However, in realistic systems the potential is {\em not}
uniform due to the screening effects.
In dc transport the actual potential 
distribution is unimportant for the evaluation of the 
conductance due to the Einstein
relation.  In contrast, the ac response is strongly sensitive
to the distribution of the potential inside the sample
\cite{Thomas,Buttik}. In its turn, the potential profile is related
via the Poisson equation to the electron density. 
Thus, the admittance has to be found self-consistently \cite{qhall}.  

Indeed, early attempts to generalize dc conductance calculations to 
the ac case (see {\em e.g.} \cite{Kramer}) within the free-electron
approach proved to be not self-consistent and failed to conserve
current. The sensitivity of the ac conductance to
different electric field configurations is illustrated in
Ref. \cite{Sablikov}. The construction of a current-conserving theory
is not an easy task. Presently, two approaches are available in the
literature. First, the self-consistent ac scattering approach
\cite{Thomas,Buttik} allowed to study the low-frequency admittance and
to express it through the scattering matrix for an arbitrary system
\cite{Buttik,Christen}. For arbitrary frequencies, the admittance is
expressed through functional derivatives of the scattering matrix with
respect to the local electric potential. Alternatively, one can employ
the methods of non-equilibrium statistical mechanics, and express the
ac response of an interacting system through the Green's functions in
a Keldysh formalism \cite{Wingreen,Bruder}. Although these methods are
quite powerful, their application to specific problems meets
substantial technical difficulties. The high frequency response of
coupled infinitely extended one-channel liquids has been investigated
by Matveev and Glazman\cite{MG}. Here we are particularly interested
in wires connected to reservoirs. Compared to infinitely extended
wires the presence of reservoirs modifies the low frequency response,
possibly up to frequencies determined by a transit time. 

Below we develop a self-consistent theory for the admittance of a
perfect ballistic wire (directed along the $x$-axis) of a length $L$
and a cross-section $S \ll L^2$ (3D) or width $W \ll L$ (2D), placed
between two reservoirs \cite{foot,Levinson}. The wire is assumed to be
shorter than all the lengths associated with inelastic scattering.
Here we treat interactions in the random phase approximation
(RPA). More specifically, we are interested in the limit where the
wire  is locally charge neutral. This is realistic in two situations:
(i) Suppose for a moment that the wire is coupled to a back-gate
through a capacitance $c$ per unit length. In the one-channel case,
this system is equivalent to the Luttinger model with short-range
interactions \cite{Blanter}; the interaction constant $g$ and the
capacitance are related by $g^2 = (1 + e^2/(\pi v_F c))^{-1}$. Here
$v_F$ is the Fermi velocity which determines the density of states
$1/\pi v_F$. The charge neutral case corresponds to the zero
capacitance limit (or $g = 0$ limit), that is, the back gate is at a
very large distance from the wire. This example also shows that the
condition of applicability of RPA $e^2 \ll v_F$ is compatible with a
small interaction constant $g$ since the capacitance can be very
small. Based on this model the results presented below are valid for
frequencies up to $v_F/R$, where $R \ll L$ is the distance between the
wire and the gate. The case of many channels can not be reduced to the
Luttinger model. (ii) In the absence of a back-gate (or for $R
\gg L$), the results presented below are valid for frequencies below
the plasma modes of a wire. For one channel these frequencies are of
order $\omega_p \sim v_F/L$ (see {\em e.g.} \cite{MG,Oparin}); for a
multi-channel wire the highest lying plasmon branch ($v_F/L$) provides
the frequency which limits the results presented below.   

We show that screening plays a crucial role for the frequency
dependence of the admittance: a wire with only one transverse
channel exhibits a monotonic frequency dependence, while a wire with
several transverse channels shows resonant structures in the
admittance, due to collective interchannel excitations in
which an accumulation of charge in one channel is locally compensated
by a charge depletion in another channel. As a consequence, the
imaginary part of the admittance changes sign as a function of
frequency. 
 
{\bf General formulation}.
The ac transport in a 1D perfect wire is determined by the following
system of equations for the local current $j$, the particle density
$\rho$, and the electric potential $\varphi/e$: 
\begin{eqnarray} 
& & j = j_p - (4\pi e)^{-1} \nabla \partial_t \varphi ; \ j_p (x)
= e \sum_a v_a \left( \rho_a^+ - \rho_a^- \right); \label{cur} \\ 
& & \Delta \varphi = -4\pi e^2 \sum_a \left(
\rho_a^+ + \rho_a^- \right); \label{Poisson} \\ 
& & -\partial_x j_p = e \partial_t 
\sum_a \left( \rho_a^+ + \rho_a^- \right); \label{contin} \\
& & \rho_a^{\pm} (x,t) = \rho_{0a}^{\pm} (x,t) \nonumber \\
& & - \int_0^{L}
\Pi_{a}^{\pm} (x,x',t-t') \varphi (x',y=0,z=0,t') dx' dt'. \label{dens} 
\end{eqnarray}
The index $a$ labels transverse channels ($1 \le a \le N_{\perp}$),
and $\rho_a^{\pm}$ denotes the excess density (with respect to the
positive background) of right/left-movers in the channel $a$, which 
depends only on the coordinate $x$. The velocities
$v_a$ of left- and right-movers in the same channel coincide, $v_a =
(\pi S \nu_a)^{-1}$. Here $\nu_a$ is the density of states in the
channel $a$; the total density of states is given by $\nu = \sum
\nu_a$. Eqs. (\ref{cur}) - (\ref{dens}) are valid within the ballistic
wire  extending from $x = 0$ to $x = L$. Furthermore, the quantities
$\rho^{\pm}_{0a}$ are ``bare'' densities of particles in the channel
$a$ {\em injected} from the left/right reservoir; the distribution function
of these particles is determined by the distribution function
of the left/right reservoir at the time of injection. Specifying 
the chemical potentials in the left/right reservoirs $\mu_L =
V(t)$, $\mu_R = 0$, we obtain  
\begin{eqnarray*} 
\rho^{\pm}_{0a} = \frac{1}{2} \nu_a \int_0^{\infty} f_{L,R} (\epsilon)
d\epsilon, 
\end{eqnarray*}
and it follows
$\rho_{0a}^+ (x,t)= \nu_a V (t -
x/v_a)/2$ and $\rho_{0a}^- = 0$. Note that for $x=L$ the bare density
of right-movers does not vanish: We do not consider in detail the
transition region between the wire and reservoirs, where the electrons
are distributed  over many quantum channels. 

Finally, $\Pi_a^{\pm} (x,x',t-t')$ is the polarization function,
responsible for the density of right/left-movers induced in the
channel $a$ at the point $x$ and time $t$ due to a potential
perturbation $\varphi$ at $x',t'$. If spatial and temporal
structure on the scale of $p_F^{-1}$ (Friedel oscillations) and
$\epsilon_F^{-1}$, respectively, can be neglected, the polarization
function quite generally has the form  
\begin{equation} \label{polar1}
\Pi_a^{\pm} (x, x',t) = (\nu_a/2) \left[ \delta(x-x')\delta(t) - 
\partial_t P_a^{\pm} (x, x',t) \right],
\end{equation}
where the function $P_a^{\pm}$ is the conditional probability to find
a right/left-moving particle in the channel $a$ at $x$ at time $t$,
if it was at $x'$ at $t'=0$. For a ballistic channel we have obviously 
\begin{equation} \label{polar2}
P_a^{\pm} (x,x',t) = \theta (t) \delta (x' \pm v_a t -x).
\end{equation}

Now we return to our system of equations. First, we note that the
Poisson equation (\ref{Poisson}) and the continuity equation
(\ref{contin}) together with the definition of the current (\ref{cur})
imply $\partial j/\partial x = 0$, {\em i.e.} the current is conserved
and does not depend on the space point. The particle current $j_p$ is
generally not conserved, and the displacement current is required for
the self-consistent treatment. However, if the system is locally
charge neutral, the case of interest here, the term $\Delta \varphi$
in Eq.(\ref{Poisson}) vanishes simultaneously with the displacement
current in Eq. (\ref{cur}). Furthermore, we note that the system  of
equations (\ref{cur})-(\ref{dens}) is excessive: $2N_{\perp} +2$
equations (\ref{Poisson}), (\ref{contin}), (\ref{dens}) contain
$2N_{\perp} + 1$ unknown fields $\varphi$ and
$\rho_a^{\pm}$. Eq. (\ref{cur}) is already a consequence of the
continuity equation and the Poisson equation. It is thus sufficient to
consider Eqs. (\ref{cur}), (\ref{Poisson}) and (\ref{dens}):  The fact
that the resulting current is position independent can be used as
simple test of consistency.  

It is convenient to Fourier-transform the
equations with respect to time, and to introduce new variables: the
full density in the channel $a$, $\rho_a = \rho_a^+ + \rho_a^-$, and
the density difference
$\zeta_a = \rho_a^+ - \rho_a^-$. 
The bare density injected by the reservoir into channel $a$ 
in response to a potential oscillation $V_{\omega}$ in that 
reservoir is $\rho_{0a}(x) = (\nu_a V_{\omega}/2) \exp (i\omega
x/v_a)$. The combined contribution of the probabilities $P^{\pm}$
gives after Fourier transform a term  
$$p_a(x) = (i\omega\nu_a/2v_a) \exp \left( i\omega \vert x \vert/v_a
\right)$$ 
in the polarization function.
It is useful also to introduce the operator $\hat Q_a$,
$$\hat Q_a g  = \nu_a g(x) + \int_0^L p_a (x - x') g(x') dx'.$$
With these abbreviations we obtain 
\begin{eqnarray}
j & = & e \sum_a v_a \zeta_a (x); \ \ \ \zeta_a (x) = \rho_{0a}
(x) \nonumber \\
& - & \int_0^L sign (x-x') p_a ( x - x') \varphi(x') dx';
\label{cur1} \\  
\sum_a \rho_a (x) & = & 0; \label{Poisson1} \\
\rho_a (x) & = & \rho_{0a} (x) - \hat Q_a \varphi (x),\label{dens1a}
\end{eqnarray}
where $\varphi(x) \equiv \varphi(x, y=0, z=0) $. Combining 
equations (\ref{Poisson1}) and (\ref{dens1a}) gives a closed 
integral equation for the potential, generated in response to the
injected density, 
\begin{eqnarray} \label{pot}
\sum_a \hat Q_a \varphi = \sum_a \rho_{0a}.   
\end{eqnarray}

Now we have to solve the equation (\ref{pot}) for the potential, and
then calculate the current from Eq. (\ref{cur1}). The admittance is
determined by $ G(\omega) = ejS/V_{\omega}.$ We re-emphasize again
that the current and thus the admittance do not depend on the space
point $x$, which is a check of the consistency of our approach. 

{\bf One channel}. For the case of one channel with the velocity $v$
the solution of Eq. (\ref{pot}) has the form 
\begin{eqnarray} \label{pot1}
\varphi(x) & = & 
V_{\omega} \left( 1 - \frac{2v}{i\omega L} \right)^{-1} 
\left( 1 - \frac{x}{L}  - \frac{v}{i\omega L} \right). 
\end{eqnarray}
For $\omega \ll v/L$ the potential is screened, and everywhere in the
wire is close to one half of the external voltage. On the other hand 
for $\omega \gg v/L$ the potential drops linearly. Eq. (\ref{pot})
describes thus the crossover from a uniform potential at low frequencies
to a uniform electric field at high frequencies. 
Calculating the
admittance, we obtain   
\begin{equation} \label{cond1}
G (\omega) = \frac{e^2}{2\pi} \left(1 - \frac{i\omega L}{2v}
\right)^{-1}. 
\end{equation} 
The imaginary part of the admittance is positive (inductive). For low 
frequencies we reproduce the behavior found previously in
Ref. \cite{Christen}, $G(\omega) = e^2N_{\perp}/\pi - i\omega E$,
with an emittance 
\begin{equation} \label{emit}
E = -\frac{e^2}{4} \nu LS.
\end{equation}

{\bf Two channels}. The Poisson equation for the potential (\ref{pot}) for 
a wire with two channels $a$ and $b$ can be solved 
by noticing that  
$$(\omega^2 + v_a^2 \partial^2/\partial x^2) \hat Q_a \varphi = \nu_a
v_a^2 \varphi''.$$ 
This implies that the potential has the form
$$\varphi(x) = \alpha + \beta x + \gamma \exp(i\xi x) + \delta
\exp(-i\xi x),$$
where the coefficients $\alpha$, $\beta$, $\gamma$ and
$\delta$ are determined by the substitution of this ansatz into 
Eq. (\ref{pot}). Interestingly,  
now, in addition to the constant and linear part, 
the potential acquires also a spatially
oscillating  part with a wave vector given by $\xi =
\omega/(v_av_b)^{1/2}$. The oscillatory structure of the potential is
also manifest in the admittance. Indeed, we find 
a non-trivial dependence of the conductance on the wave vector 
$\xi$ and the density of states ratio $\eta = (v_a/v_b)^{1/2}$,
\end{multicols}
\widetext
\vspace*{-0.2truein} \noindent \hrulefill \hspace*{3.6truein}
\begin{equation} \label{cond2}
G(\omega) = \frac{e^2}{2\pi} (\eta^2 + 1) \left( - \frac{i\xi
L\eta}{2} + 
\frac{(\eta^3+1)(\eta+1) - (\eta^3-1)(\eta - 1)\exp(i\xi
L)}{(\eta+1)^2 - (\eta -1)^2\exp(i\xi L)} \right)^{-1} \nonumber
\end{equation}  
\hspace*{3.6truein}\noindent \hrulefill 
\begin{multicols}{2} 
\noindent The conductance is symmetric with
respect to the replacement $\eta \to \eta^{-1}$, as it must be. For
low frequencies we reproduce again the static quantized conductance 
$e^2/\pi$ and the emittance (\ref{emit}). The real part is strictly positive,
although now both real and imaginary part exhibit oscillations on
top of the monotonic behavior found for the single channel wire.
In the limit $v_a = v_b$, 
which corresponds to a spin degenerate one-channel 
conductor, these oscillations vanish. The behavior
of the admittance as a function of frequency for different values of
the parameter $\eta$ is shown in Fig.~1. Note that the imaginary part
may change sign in the vicinity of the points $\omega \sim 2\pi n
(v_a v_b)^{1/2} /L$ (see below), if $\eta$ is large enough.
A large $\eta$ occurs for Fermi energies just above
the threshold of the second conductance channel. 

The oscillatory structure of the admittance can be understood by
investigating the 
poles of the admittance. 
In the limiting case $\eta
\ll 1$, we obtain for the spectrum of collective modes, 
$$\omega = \frac{(v_av_b)^{1/2}}{L} \left[ 2\pi n (1 - \eta^2)  - 2 i
\eta \right], \ \ \ \ n \in Z.$$ 
The purely imaginary eigenvalue with $n=0$, as for one channel,
corresponds to charge relaxation between the wire and the reservoirs 
via ballistic motion, while the resonances
for $n \ne 0$ correspond to nearly neutral interchannel
excitations. These modes are essentially standing waves induced in both
channels simultaneously but with densities of opposite signs, $\rho_a
= -\rho_b \propto \exp(i\xi x) - \exp(i\xi(L-x))$. The decay of these
excitations, $\mbox{Im} \ \omega = -2v_a/L$, is determined by the
carriers in the channel with the lower velocity. Interestingly, 
we find that due to the coupling to reservoirs all the collective modes
are damped with a relaxation constant which is the larger the 
shorter the wire is. 

{\bf Many channels}. For $N_{\perp} > 1$ channels the potential has the
form 
$$\varphi(x) = \alpha + \beta x + \sum_{i=1}^{N_{\perp}-1} \left[
\gamma_i \exp(i\omega\lambda_i x) + \delta_i \exp(-i\omega\lambda_i
x) \right],$$  
and the positive quantities $\lambda_i$ are solutions of the equation
\begin{equation} 
\sum_a \frac{v_a}{1 - v_a^2 \lambda^2} = 0.
\label{ya}
\end{equation}
For arbitrary $N_{\perp}$ and $\omega$, further analytical progress
is hard, but the problem can be solved for
low frequencies $\omega \ll \min (v_a/L)$. We obtain 
\begin{equation} \label{cond3}
G(\omega) = \frac{e^2 N_{\perp}}{2\pi} + i\omega \frac{e^2}{4} \nu SL
-\omega^{2} \frac{e^2\pi}{8} (SL)^2 \sum_a \nu_a^2 + \dots \ \ .
\end{equation} 
 
In conclusion, we investigated the admittance of a perfect ballistic
wire in the frequency range below $v_F/L$. We showed that the
screening effects are very important for the admittance, and
provide the current conservation. For one channel, the admittance is a
monotonic function of frequency, whereas for two or more channels it
contains also oscillatory components due to the density redistribution
between different channels. In particular, the imaginary part of the
admittance is generally positive (inductive-like), but can change sign 
and become capacitive in the vicinity of resonances due to the
interchannel excitations. The wire exhibits resonances due to damped
collective modes. The resonant effects predicted can be measured
experimentally; the collective mode frequencies depend on the relative
electron concentration in the different channels and near the
threshold for a new quantum channel, where the resonances are most
pronounced, can be made very small. 

We thank L.~I.~Glazman, D.~E.~Khmel'nitskii, and K.~A.~Matveev for
useful discussions. The work was supported by the Swiss National
Science Foundation.  

\begin{figure}
\narrowtext
\centerline{\psfig{figure=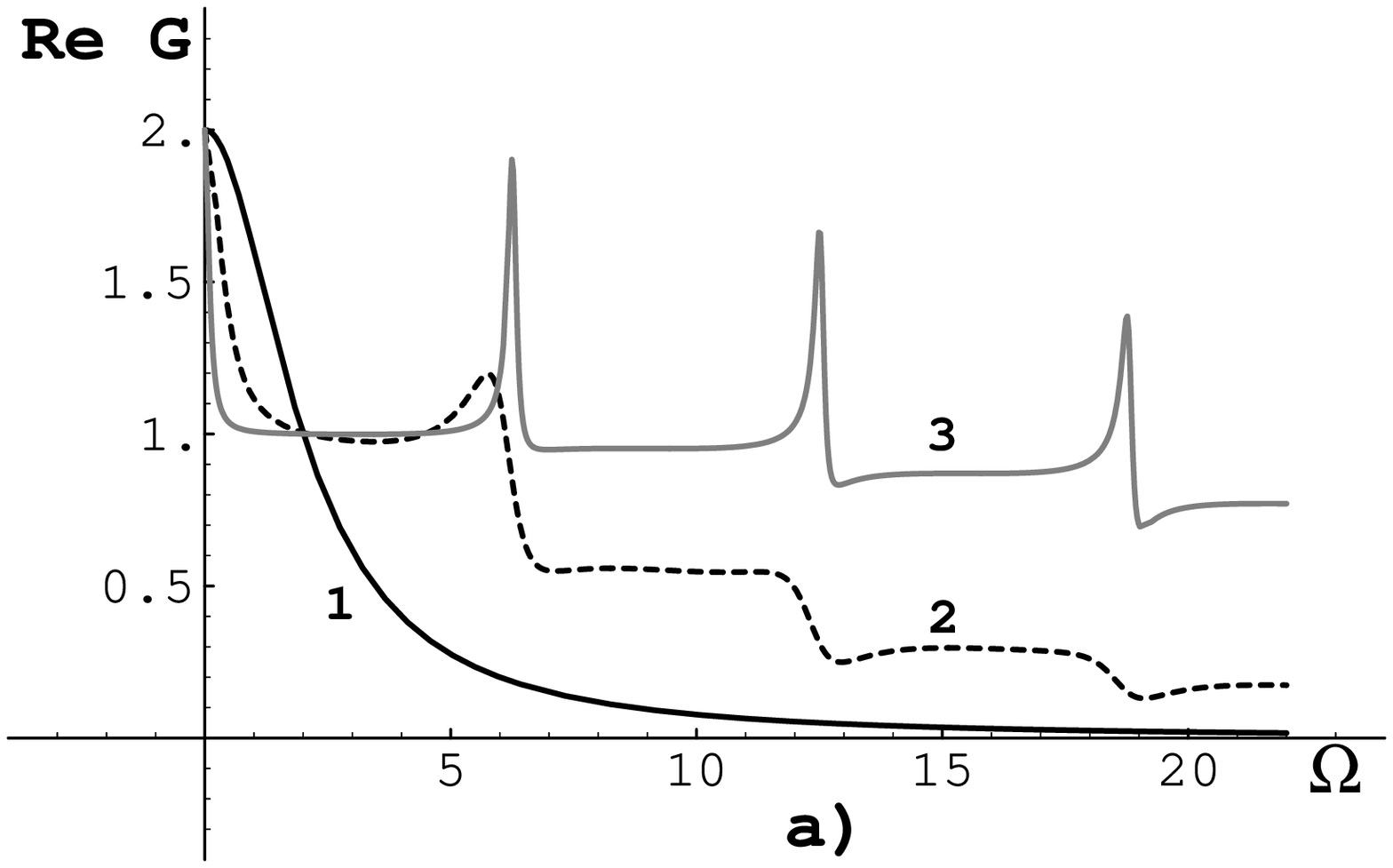,width=7.cm}}
\centerline{\psfig{figure=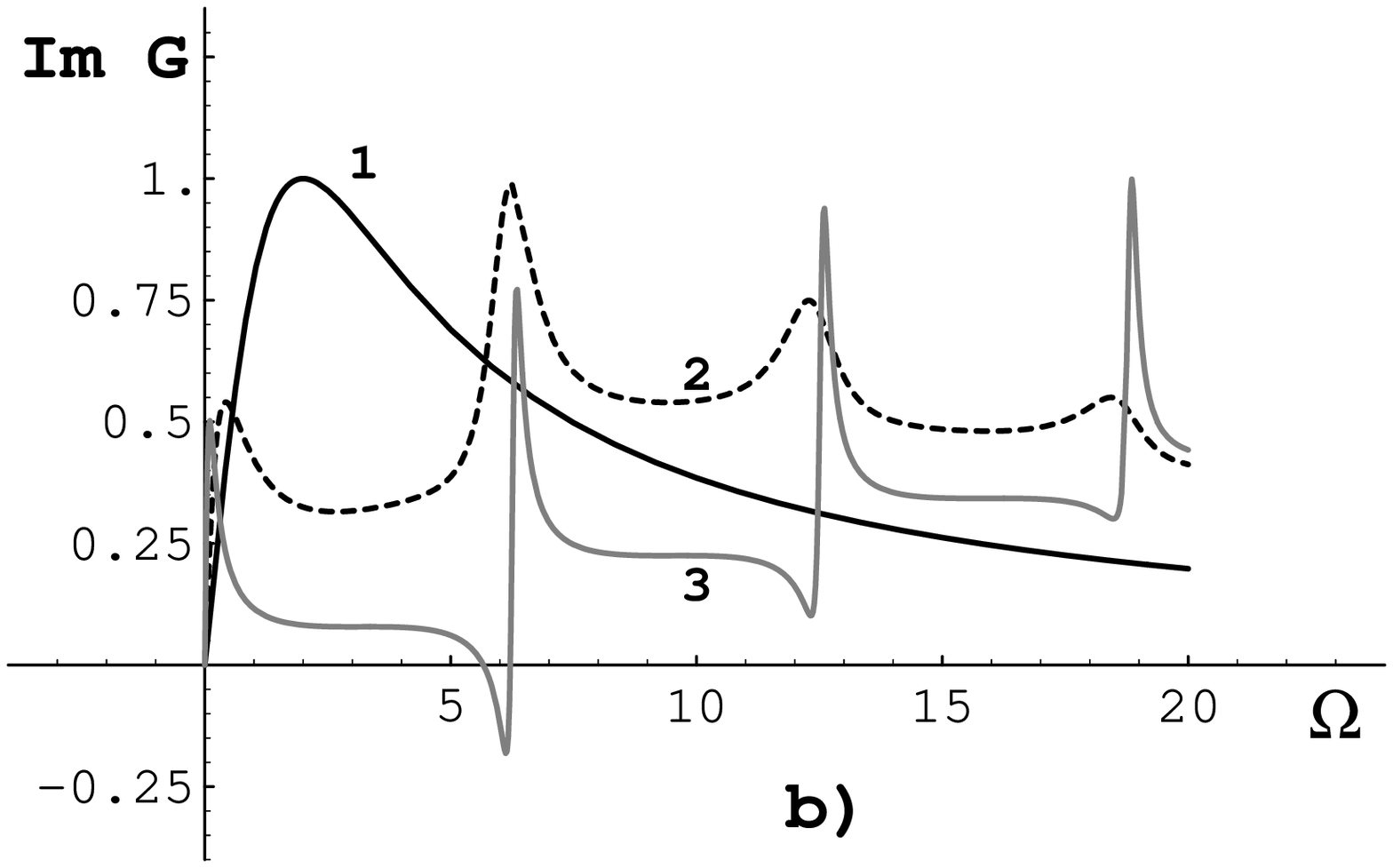,width=7.cm}}
\vspace{0.3cm}
\caption{Real (a) and imaginary (b) parts of the conductance in units
of $e^2/2\pi$ for the two-channel case as a function of the parameter
$\Omega = \omega L/(v_a v_b)^{1/2}$; the parameter $\eta =
(v_a/v_b)^{1/2}$ is equal to 1 (Curve 1), 5 (2) and 20 (3). Resonances
are due to the interchannel excitations.}
\end{figure}

\end{multicols} 

\begin{thebibliography}{99}
\bibitem{foot} In the following we use a notation 
appropriate for 3D; in a 2D geometry $S$ should be replaced by
$W$. The number of transverse channels is of order $N_{\perp} \sim
p_F^2S$ (3D) or $N_{\perp} \sim p_FW$ (2D). We consider the case of
zero temperature, and use the system of units with $\hbar = k_B = 1$.
\end{thebibliography}
\end{document}